\documentclass[preprint,onecolumn,floatfix,superscriptaddress,amsmath,amsfonts,amssymb,aps]{revtex4}

\usepackage{graphicx}
\usepackage{dcolumn}
\usepackage{bm}
\usepackage[normalem]{ulem}
\usepackage{color}
\usepackage{amssymb}
\usepackage{amstext}

\begin{document}


\title{Localization effects in the disordered Ta interlayer of multilayer Ta--FeNi films: Evidence from dc transport and spectroscopic ellipsometry study}

\author{N. N. Kovaleva}
\email{kovalevann@lebedev.ru}
\affiliation{Institute of Physics, Academy of Sciences of the Czech Republic, Na Slovance 2, 18221 Prague, Czech Republic}
\affiliation{Lebedev Physical Institute, Russian Academy of Sciences, Leninsky prospect 53, 119991 Moscow, Russia}
\affiliation{Department of Physics, Loughborough University,LE11 3TU Loughborough, United Kingdom} 
\author{\mbox{D. Chvostova}}
\affiliation{Institute of Physics, Academy of Sciences of the Czech Republic, Na Slovance 2, 18221 Prague, Czech Republic}
\author{O. Pacherova}
\author{L.~ Fekete}
\affiliation{Institute of Physics, Academy of Sciences of the Czech Republic, Na Slovance 2, 18221 Prague, Czech Republic}
\author{K. I. Kugel}
\affiliation{Institute for Theoretical and Applied Electrodynamics, Russian Academy of Sciences, 125412 Moscow, Russia}
\author{F. A. Pudonin}
\affiliation{Lebedev Physical Institute, Russian Academy of Sciences, Leninsky prospect 53, 119991 Moscow, Russia}
\author{A. Dejneka}
\affiliation{Institute of Physics, Academy of Sciences of the Czech Republic, Na Slovance 2, 18221 Prague, Czech Republic}

\date{\today}

\begin{abstract}
Using dc transport and wide-band spectroscopic ellipsometry techniques, we study localization effects in the  disordered metallic Ta interlayer 
of different thickness in the multilayer films (MLFs) (Ta -- FeNi)$_{\rm N}$ grown by rf sputtering deposition. In the grown MLFs, the FeNi layer was 0.52\,nm thick, while the Ta layer thickness varied between 1.2 and 4.6\,nm. The Ta layer dielectric function was extracted from the Drude-Lorentz 
simulation. The dc transport study of the MLFs implies non-metallic 
(${\rm d}\rho/{\rm d}T$\,$<$\,0) behavior, with negative temperature 
coefficient of resistivity (TCR). The TCR absolute value increases 
upon increasing the Ta interlayer thickness, indicating enhanced 
electron localization. With that, the free charge carrier Drude response decreases. Moreover, the pronounced changes occur at the extended spectral range, involving the higher-energy Lorentz bands. 
\mbox{The Drude} dc conductivity drops below the weak localization 
limit for the thick Ta layer. The global band structure reconstruction may indicate the formation of a nearly localized many-body electron state. 
\end{abstract}

\pacs{Valid PACS appear here}
\maketitle
Under the condition of increased disorder, scattering by static 
structural defects may occur more frequently than inelastic 
scattering by phonons in the appropriate temperature range. 
This condition may be fulfilled in high-resistivity alloys, 
where the mean free path of conduction electrons in the process of 
elastic scattering at the static defects is of the order of magnitude of interatomic distance $l_e$\,$\sim$\,$a$\,$\sim$\,$k_{\rm F}^{-1}$ (where $k_{\rm F}$ is the Fermi wavenumber). It was proposed that due to this condition, weak localization correction \cite{Gorkov} may take place in disordered metals  and determine non-metallic character of their dc transport and negative temperature coefficient of resistivity (TCR) \cite{Mooij} 
till room temperature \cite{Tsuei,Gantmakher}.

$\beta$-Ta\,films are known to possess negative TCR \cite{Read}. Recently, using dc transport and wide-band spectroscopic ellipsometry techniques, we have studied $\beta$-Ta films grown by rf sputtering deposition \cite{Kovaleva_APL}. We found that the temperature variation of their dc transport, $\rho(T)=\rho_0\left[ 1+\alpha_0(T-T_0)\right]$, shows non-metallic behavior (${\rm d}\rho/{\rm d}T$\,$<$\,0) with negative TCR ($\alpha_0$). The determined ($\alpha_0$,\,$\rho_0$) values well fit the range of the Mooij plot for disordered or amorphous metals having negative TCR \cite{Mooij,Tsuei} and show a similar trend. 
This implies that the physics of the studied $\beta$-Ta films is driven by static disorder. An additional spectroscopic ellipsometry probe showed that with increasing the TCR absolute value, specifying an elevated degree of disorder, the free charge carrier Drude response decreases. Moreover, it was found that the pronounced changes occur in the extended spectral range, involving also the high-energy Lorentz bands, related to the electron correlation effects accompanying the weak localization \cite{Kovaleva_APL}.

Here, we would like to explore the elaborated approach \cite{Kovaleva_APL}, which combines dc transport and wide-band spectroscopic ellipsometry techniques, to gain insights into the localization phenomena for disordered metallic systems, driven by additional magnetic disorder. Anderson localization is expected under conditions of strong magnetic disorder. For example, sputtered ultrathin Fe$_{21}$Ni$_{79}$ permalloy films show complex disordered structures composed of single-domain ferromagnetic (FM) nanoislands, 
comprising many atomic moments $\sim$10$^3$--10$^5$\,$\mu_{\rm B}$, where $\mu_{\rm B}$ is the Bohr magneton. Therefore, a study of multilayer composite metallic films, incorporating FeNi nanoisland layers, provides an ideal platform for exploring the Anderson localization mechanism in two dimensions. By controlling the thickness of the metallic layer, one can tune the critical distance from the magnetic nanoisland layer for the onset of the Anderson localization, which can lead to a many-body localized state \cite{Basko}. 
We developed an experimentally relevant system of multilayer composite metallic Ta films (Ta\,--\,FeNi)$_{\rm N}$, incorporating FeNi layers, represented by inhomogeneously distributed flat FM nanoislands \cite{Sherstnev,Stupakov}.
\begin{figure}[b]\vspace{-0.8em}
        \includegraphics[width=\columnwidth]{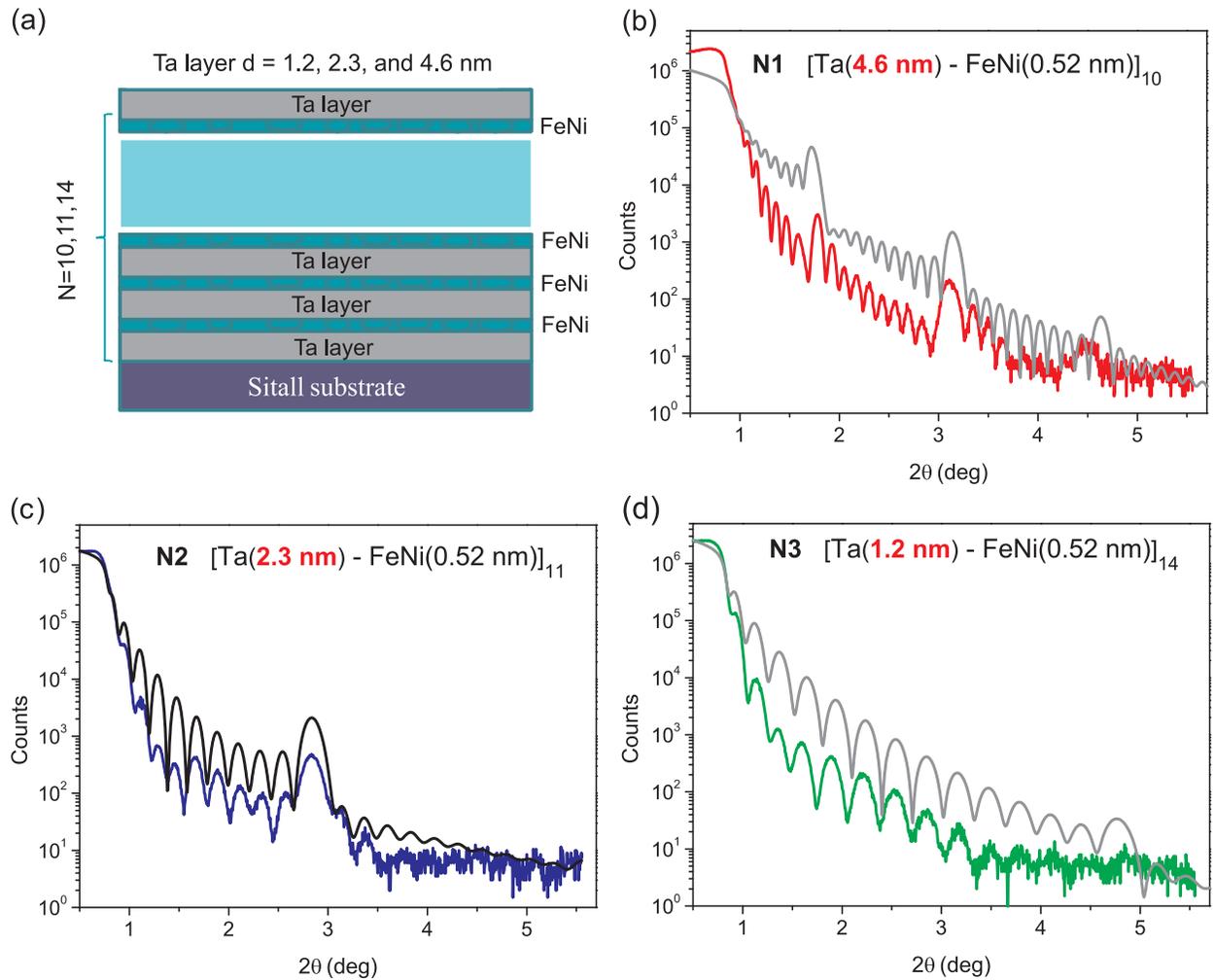}\vspace{-0.3em}
\caption{(a) Schematics of the grown MLFs (Ta--FeNi)$_{\rm N}$. The nominal thickness of the FeNi layer was 0.52\,nm. (b)--(d) X-ray reflectivity of the MLF samples N1, N2, and N3, respectively. Gray lines show the result of computational simulations.}
\vspace{-0.6cm}
\label{SLscheme}
\end{figure}
The multilayer films (MLFs) (Ta\,--\,FeNi)$_{\rm N}$ were grown by rf sputtering deposition from 99.95 \% pure Ta and Fe$_{21}$Ni$_{79}$ targets on an insulating glass Sitall substrate. The vacuum chamber was subjected to annealing at the temperature of 200$^\circ$C for 1 h. The chamber base pressure before beginning the rf sputtering was about 2\,$\times$\,10$^{-6}$ Torr. The background Ar pressure was 6$\times$10$^{-4}$ Torr, and the actual substrate temperature was about 80 $^\circ$C. We used the Sitall substrates with typical sizes of 15$\times$5$\times$0.6 mm$^3$. The Ta and FeNi layer nominal thickness was determined by the deposition time defined by the film deposition rate. 

In the grown MLF samples N1, N2, and N3 [schematically illustrated by Fig.\,\ref{SLscheme}(a)], the FeNi layer nominal thickness was 0.52\,nm, and the thickness of the Ta interlayer was 4.6, 2.3, and 1.2 nm, where the number of layers is N = 10, 11, and 14, respectively. Below the percolation transition  at the critical thickness $d_c$\,$\simeq$\,1.5\,--\,1.8 nm, the FeNi layer has a nanoisland structure \cite{Stupakov,Sherstnev}. The lateral sizes of the islands are 5\,--\,30 nm, and the distance between them is 1\,--\,5 nm. It was shown that in the multilayer structures of periodically alternating layers of nanoislands and continuous layers, they do not mix \cite{Pudonin}. This is verified by the X-ray reflectivity showing three Bragg peaks in evidence of a good periodicity. We would also anticipate that the 1.2\,nm Ta layer in the sample N3 has a nanoisland structure. Then, the sample N3 is composed of periodically alternating nanoisland FeNi and Ta layers. It was shown that in periodic nanoisland multilayer structures, the island layer mixing is insignificant \cite{Pudonin}. However, since nanoislands of neighboring layers are practically in contact with each other, the nanoisland metallic multilayer structure will be percolating. We would also expect that the thinner Ta layer grown at the same rf sputtering conditions will possess a higher degree of disorder. Indeed, elevation of the MLF surface temperature caused by rf discharge leads to the annealing effect and promotes the effective mechanism of defect annihilation upon increasing the Ta layer thickness. By contrast, we suppose that the concentration of oxygen defects, which can be caused by the presence of oxygen at a background level of 10$^{-8}$\,--\,10$^{-9}$ mbar in Ar gas, will be much less thickness dependent, as its absorption occurs permanently.\,The grown MLFs were capped {\it in situ} with a 2.1\,nm thick Al$_2$O$_3$ layer.

X-ray reflectometry was used to characterize layered structures of the grown MLF samples (Ta\,--\,FeNi)$_{\rm N}$. 
We utilized Bede 200 Goniometer operating at 55 kV and 300 mA with Cu K$\alpha$ radiation produced by an X-Ray Generator with rotating anode Rigaku RU 300. For the MLF structure analysis, we used the Bruker Diffrac plus Leptos computational simulation, based on the theory of X-ray reflectometry. In the simulation, all layers in the MLF structure were considered to be homogenous pure
materials. Ta and FeNi components were considered to be identical through
the whole MLF structure. The best fit was searched by changing the thicknesses and densities of the Ta and FeNi layers, and some roughness of
interfaces was also taken in account. In Figs.\,\ref{SLscheme}(b)--\ref{SLscheme}(d) we show the X-ray reflectivity data for the studied MLF samples N1, N2, and N3 and the result of computational simulation. The basic features, obtained from the simulation process, like the parameter of periodicity, the whole thickness of the set of layers, and the layer densities, were consistent with the structure and content of the grown MLFs. The MLF N1 shows three Bragg peaks in evidence of a good periodicity [see Fig.\,\ref{SLscheme}(b)]. Missing Bragg peaks for the MLF N3 [see Fig.\,\ref{SLscheme}(d)], where the Ta layer is non-continuous, can be related to the small ratio of the FeNi and Ta layer densities. In this case, it is possible that X-rays do not detect the superlattice structure. The estimated thickness of the Ta (FeNi) layer of 4.86 (0.96), 2.68 (0.5), and 1.37\,(0.48) nm is in good agreement with the nominal Ta (FeNi) thickness values in the MLFs N1, N2, and N3, respectively. The estimated density of the Ta (nanoisland FeNi) layer was 14.2 (4.1), 10.2 (3.4), and 11.6 (5.8) g/cm$^3$. The estimates are in satisfactory agreement with the bulk Ta density of 16.6 g/cm$^3$. It is also reasonable to assume that the estimated 
density of the nanoisland FeNi layer is less than the bulk FeNi density of 8.62 g/cm$^3$. The estimated roughness for the Ta and FeNi interfaces in the MLFs was about 0.6\,nm. Some features, however, were not explained well by the simulation, most probably due to inhomogeneous character of the FeNi layer.  

\begin{figure}[b]\vspace{-0.8em}
        \includegraphics[width=120mm]{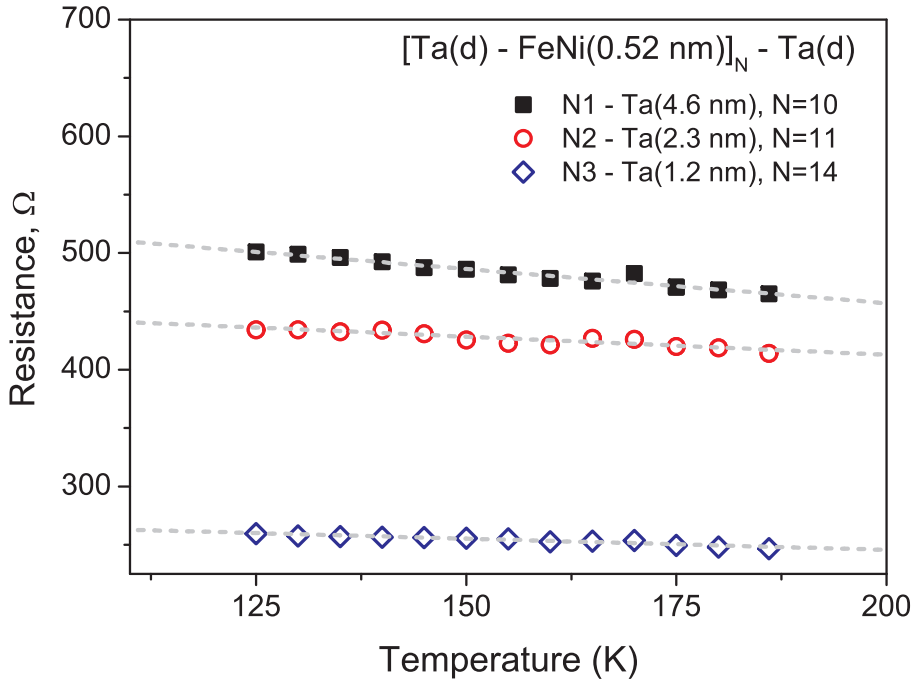}\vspace{-0.3em}
\caption{Temperature dependence of the sheet dc resistance of the MLFs N1, N2, and N3. Dashed lines correspond to the linear approximation 
(see the text).}
\vspace{-0.6cm}
\label{FigresSL}
\end{figure}
The objective of our present dc transport measurements was precisely to 
study how the negative TCR regime of $\beta$-Ta layers is affected by additional structural and/or magnetic disorder in the MLF samples (Ta\,--\,FeNi)$_{\rm N}$--Ta/Sitall N1, N2, and N3. Sheet dc resistance of the MLFs samples (Ta\,--\,FeNi)$_{\rm N}$--Ta/Sitall N1, N2, and N3 was measured on cooling from 180 K down to 125\,K, using the four-point probe home-made system. For that, the MLF samples were mounted on a cold finger of the helium cryostat with an optimized temperature controller supporting a temperature stability of $\pm$\,0.1$^\circ$C. The used multichannel electrical circuit allowed us to measure the temperature dependence of the sheet dc resistance simultaneously on several samples. Figure\,\ref{FigresSL} shows the temperature dependence of the sheet dc resistance of the MLF samples N1, N2, and N3, having different thickness of the Ta layer:\ 4.6, 2.3, and 1.2\,nm, respectively. One can see that the measured temperature dependence demonstrates non-metallic (${\rm d}\rho/{\rm d}T$\,$<$\,0) behavior. Below 180 K, their sheet dc resistance is well approximated by the linear temperature dependence having a negative slope: $R_{\rm N1}(T)$\,$\simeq$--\,0.585($\pm$0.037)\,$T$\,+\,574($\pm$6)\,$\Omega$, $R_{\rm N2}(T)$\,$\simeq$--\,0.311($\pm$0.038)\,$T$\,+\,475($\pm$6)\,$\Omega$, and $R_{\rm N3}(T)$\,$\simeq$--\,0.192($\pm$0.016)\,$T$\,+\,284($\pm$3)\,$\Omega$ (see Fig.\,\ref{FigresSL}). The negative slope decreases in the investigated series of the MLF samples N1, N2, and N3 with decreasing the Ta layer thickness. This is in contrast to the recently studied single layer $\beta$-Ta films, grown on the Sitall substrate at the same rf sputtering conditions \cite{Kovaleva_APL}. There, the absolute value of negative TCR increases with the increasing degree of disorder in the thinner $\beta$-Ta single layer films. 
The TCR behavior in the studied MLFs can be associated with the presence of inhomogeneous FM nanoisland FeNi layers, which have profound effects on the conductivity properties of the Ta layer.

Complex dielectric function spectra of the MLFs (Ta\,--\,FeNi)$_{\rm N}$\,--\,Ta/Sitall [schematically illustrated in Fig.\,\ref{SLscheme}(a)] were investigated in the wide photon energy range 0.8\,--\,8.5\,eV with a J.A. Woollam VUV-VASE Gen II spectroscopic ellipsometer. The ellipsometry measurements were performed at two incident angles of 65$^\circ$ and 70$^\circ$ at room temperature. Additionally, the corresponding ellipsometry measurements were performed on the blank Sitall substrate. For each angle of incidence, the raw experimental data are represented by real values of the ellipsometric angles, $\Psi(\omega)$ and $\Delta(\omega)$. These values are defined through the complex Fresnel reflection coefficients for light-polarized parallel $r_p$ and perpendicular $r_s$ to the plane of incidence as 
${\rm tan}\,\Psi\,e^{i\Delta}=\frac{r_p}{r_s}$. 
The measured ellipsometric angles, $\Psi(\omega)$ and $\Delta(\omega)$, were simulated by multilayer models using the J.A. Woollam VASE software (see more details in the supplementary materials). 
\begin{figure}[b]\vspace{-0.8em}
        \includegraphics[width=120mm]{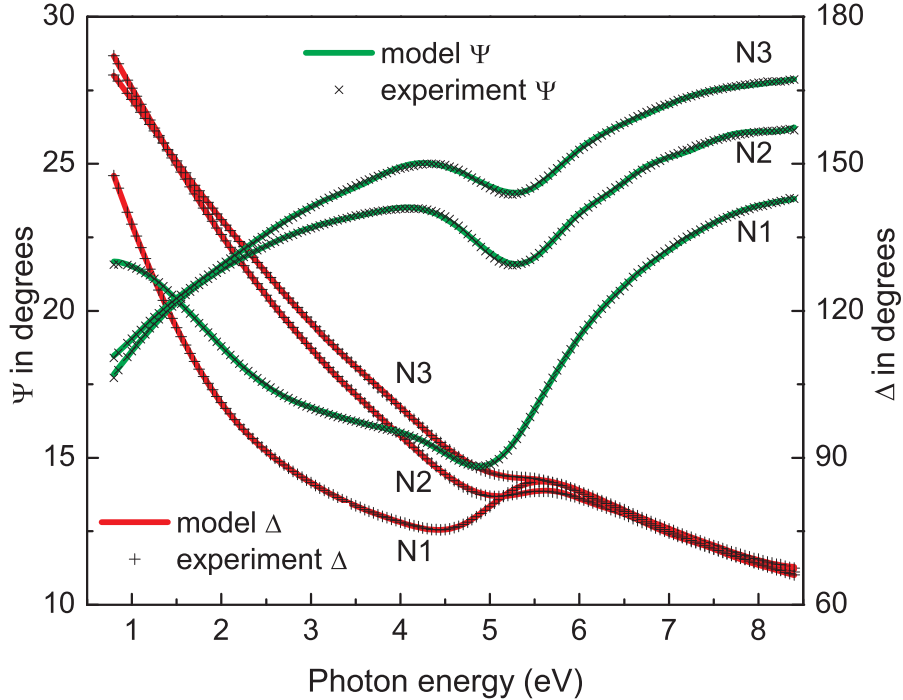}\vspace{-0.3em}
\caption{The ellipsometric angles, $\Psi(\omega)$ and $\Delta(\omega)$, measured at the angle of incidence of 70$^\circ$ and the fitting results by the Drude-Lorentz model [Eq.\,(\ref{DrLor})] for the MLFs N1, N2, and N3.}
\vspace{-0.6cm}
\label{psiDeltaTa}
\end{figure}

The complex dielectric function response, $\tilde \varepsilon(\omega)=\varepsilon_1(\omega)+{\rm i}\varepsilon_2(\omega)$, of each layer was modelled by a Drude term, responsible for free charge carrier response and contributions from Lorentz oscillators
\begin{eqnarray}
\varepsilon=\epsilon_{\infty}-\frac{\hbar^2}{\varepsilon_0\rho\left( \tau E^2+{\rm i}\hbar E \right)}+\sum_j\frac{S_jE^2_j}{E_j^2-E^2-{\rm i}E\gamma_j}. \label{DrLor}
\end{eqnarray}
Here, $E\equiv\hbar\omega$, the Planck's constant $\hbar$ and the vacuum dielectric constant $\varepsilon_0$ are the physical constants, and $\epsilon_{\infty}$ is the core contribution to the dielectric function. The zero-frequency resistivity $\rho$ 
and the mean scattering time $\tau$ are the variable
parameters in fitting the Drude term. In addition, $E_j$, $\gamma_j$, and $S_j$ are the fitted parameters of the peak energy, full width at 
half maximum, and oscillator strength of the $j^{th}$ Lorentz oscillator, respectively.

\begin{figure}[b]\vspace{-0.8em}
        \includegraphics[width=\columnwidth]{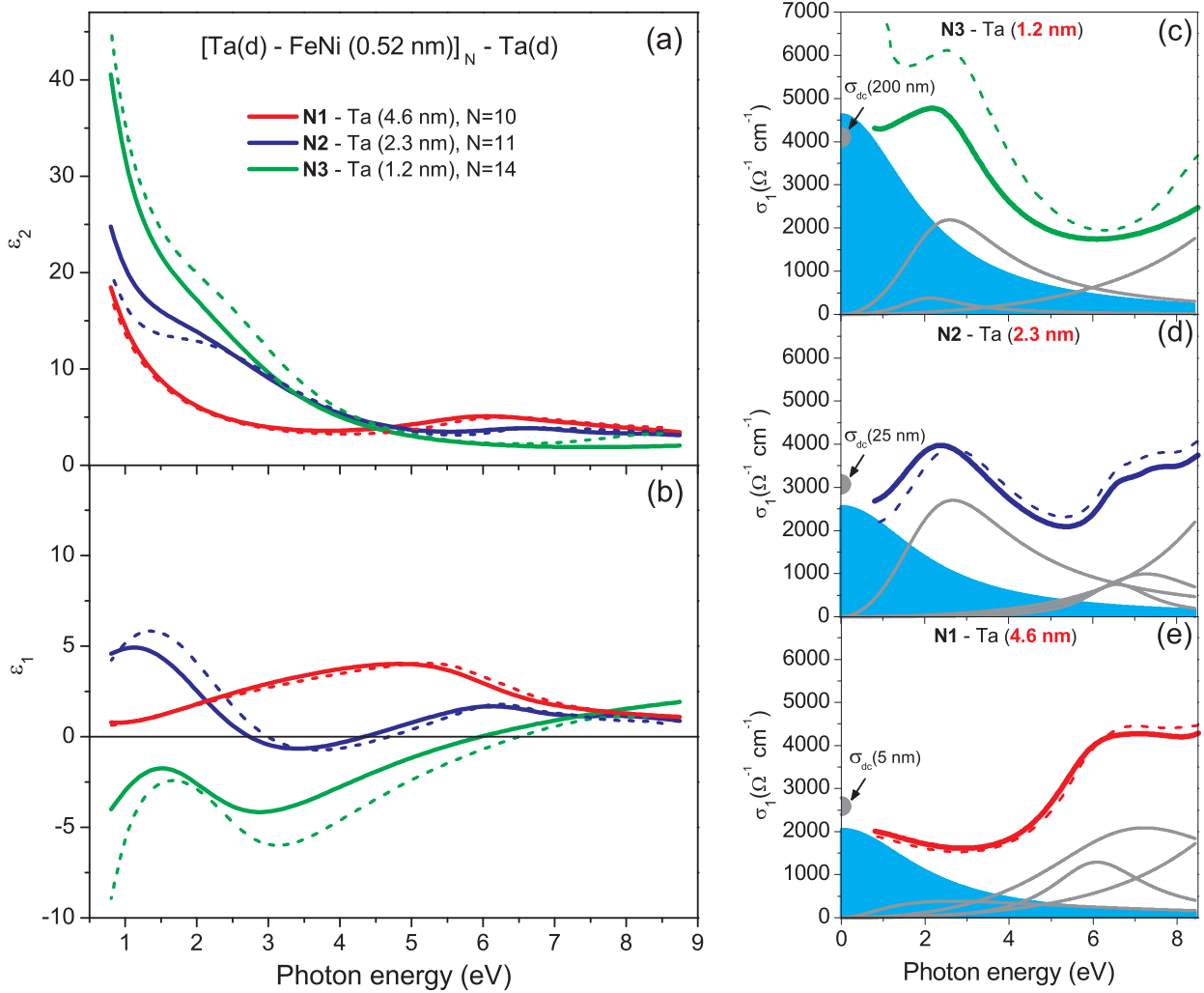}\vspace{-0.3em}
\caption{(a) $\varepsilon_2(\omega)$ and (b) $\varepsilon_1(\omega)$ dielectric function spectra of the Ta layer of different thickness in N1, N2, and N3, shown by solid red, blue, and green curves, respectively. The simulation for N1 and N2 with 0.6 nm roughness and EMA for N3 are shown by the dashed curves. (c)--(e) The Ta layer optical conductivity, $\sigma_1(\omega)=\frac{1}{4\pi}\omega \varepsilon_2(\omega)$. The Drude term (cyan shaded area) and Lorentz oscillators (solid gray lines) contributions.}
\vspace{-0.6cm}
\label{SW}
\end{figure}

The MLFs N1, N2, and N3 were simultaneously fitted by using the Drude-Lorentz model [Eq.\,(\ref{DrLor})]. The multisample fitting procedure was used to minimalize the number of free parameters. This method is commonly used for a set of samples having an identical dielectric function of incorporated layers. The VASE software makes this procedure possible. In the simulation of the ellipsometry data, the layers having a nanoisland structure (the FeNi layer in the MLFs N1, N2, and N3 and the Ta layer in the sample N3) were represented by effective dielectric function. 
The FeNi layer effective dielectric function was represented by one dispersion model in all MLFs, including three identical Lorentz oscillators. The Lorentz parameters were fitted dependently (simultaneously). During the fit, the FeNi layer thickness was fixed at the nominal thickness value of 0.52\,nm. The Ta layer in each MLF was described by a different dispersion model, including the Drude term and three or four Lorentz oscillators; however, the parameters were fitted independently. The high quality of the fit is demonstrated in Fig.\,\ref{psiDeltaTa}, where we present the measured ellipsometric angles, $\Psi(\omega)$ and $\Delta(\omega)$, along with the fitting results. 
The simulation with the incorporated intermix layer (an effective medium with the layer thickness equivalent to the interface roughness) or interface roughness does not improve the fit. The thickness of the Ta layer estimated from the simulation was 4.6\,$\pm$\,0.1\,nm, 2.10\,$\pm$\,0.07\,nm, and 1.27\,$\pm$\,0.07\,nm, in good agreement with the respective nominal thickness values in the MLFs N1, N2, and N3. This means that the Ta\,--\,FeNi interface roughness  (of about 0.6\,nm, as estimated from the present X-ray reflectometry study) is essentially incorporated in the effective dielectric function of the nanoisland FeNi layer. From the simulation, the complex dielectric function of the Ta layer of different thickness in the investigated MLFs (Ta\,--\,FeNi)$_{\rm N}$\,--\,Ta/Sitall was extracted. The extracted effective dielectric function of the porous Ta layer in the sample N3 was subsequently simulated in the effective medium approximation (EMA) by Ta\,:\,24\% air voids. The obtained Ta intralayer $\varepsilon_2(\omega)$ and $\varepsilon_1(\omega)$ are displayed in Fig.\,\ref{SW}(a,b), respectively. One can see that the Ta intralayer $\varepsilon_2(\omega)$ dramatically decreases at low probed photon energies  upon increasing the Ta layer thickness in the MLFs. At the same time, one can notice that the Ta intralayer $\varepsilon_1(\omega)$ of the MLFs N2 and N3 displays down-turn to negative values at the lowest probed photon energies, indicating a trend towards metallic-like behavior. This is in contrast to the $\varepsilon_1(\omega)$ behavior for the single layer $\beta$-Ta films of different thickness, grown at the same rf sputtering conditions, where it showed the peculiar non-metallic-like behavior, with up-turn to positive values \cite{Kovaleva_APL}. This behavior was associated with the presence of the pronounced Lorentz band around 2\,eV, being a signature of the important role of electron correlations \cite{Kovaleva_APL}.      

Figures\,\ref{SW}(c)--\ref{SW}(e) display evolution of the Ta intralayer optical conductivity, $\sigma_1(\omega)=\frac{1}{4\pi}\omega \varepsilon_2(\omega)$, upon increasing the Ta layer thickness in the MLFs (Ta\,--\,FeNi)$_{\rm N}$. Here, the contributions from the Drude term and Lorentz oscillators obtained from the Drude-Lorentz modeling [Eq.\,(\ref{DrLor})] of the complex dielectric function of the MLFs [shown in Figs.\,\ref{SW}(a) and \ref{SW}(b)] are explicitly demonstrated (as in Ref.\,\cite{Boris}). From Figs.\,\ref{SW}(c)--\ref{SW}(e), one can notice that the Drude term notably decreases upon increasing the Ta layer thickness (1.2\,nm\,$\rightarrow$\,2.3\,nm\,$\rightarrow$\,4.6\,nm). Moreover, the optical conductivity changes are extended to the higher photon energies, involving the low-energy Lorentz bands (around 2\,--\,3\,eV) and the high-energy Lorentz bands (around 6\,--\,8\,eV). A very similar trend was observed with the increasing degree of disorder in the single layer $\beta$-Ta films upon decreasing the film thickness from 200 to 5\,nm \cite{Kovaleva_APL}. The estimated values of the Drude dc limit $\sigma^{opt}_{\omega\rightarrow0}$ of the Ta intralayer $\sim$\,2550 (2.3\,nm) and $\sim$\,2050 (4.6\,nm) $\Omega^{-1}\cdot$\,cm$^{-1}$ are somewhat less than those of the single layer 5--70\,nm thick $\beta$-Ta films [see Figs.\,\ref{SW}(d) and \ref{SW}(e)] \cite{Kovaleva_metals}. This may reflect some drawbacks of the present ellipsometry model. However, the main trend [shown in Figs.\,\ref{SW}(a)--\ref{SW}(e)] is reproducible for sensible values of the incorporated intermix layers or interface roughness. We note that the effective Drude dc conductivity of the 1.2\,nm thick Ta layer of $\sim$\,4640\,$\Omega^{-1}\cdot$\,cm$^{-1}$ is higher than that of the well annealed 200\,nm thick $\beta$-Ta film (of $\sim$\,4100 $\Omega^{-1}\cdot$\,cm$^{-1}$). The EMA simulation of the porous Ta layer in the sample N3 [Figs.\,\ref{SW}(a) and \ref{SW}(c)] suggests that the conductivity difference is positive and constitutes more than 12\% (it is also possible that there is an additional contribution from the narrow Drude peak at low photon energies). For comparison, the positive magnetoconductivity effect determined from far-IR reflectivity spectra of the colossal magnetoresistive (CMR) compound La$_{0.67}$Ca$_{0.33}$MnO$_3$ in an external magnetic field of 16 T is 16\% at T=307 K \cite{Boris2}. 
However, for the 4.6\,nm thick Ta layer, the Drude dc conductivity drops well below the weak localization limit for disordered metals 1/$\rho_0^*$\,$\simeq$\,3300\,$\Omega^{-1}\cdot$\,cm$^{-1}$, where the resistivity $\rho_0^*=\frac{\hbar k_{\rm F}}{ne^2l_e}=\frac{\hbar}{e^2}\frac{1}{k_{\rm F}}$\,$\simeq$\,300\,$\mu \Omega\cdot$\,cm corresponds to an average mean free path $l_e$$\sim$$a$$\sim$$k_{\rm F}^{-1}$. From Fig.\,\ref{SW}(e) one can follow the pronounced increase of the Lorentz band around 6\,--\,8\,eV, while the Lorentz band at 2 eV becomes essentially suppressed. We suggest that this may indicate the transition to a many-body localized state, where the additional polaronic-like contribution appears (with the binding energy 2$E_b$\,$\sim$\,6\,--\,8\,eV).

In summary, the dc transport study of the MLFs (Ta--FeNi)$_{\rm N}$ implies non-metallic (${\rm d}\rho/{\rm d}T$\,$<$\,0) behavior with negative TCR. The TCR absolute value increases upon increasing the Ta layer thickness, implying enhanced localization effects. The extracted Ta intralayer optical conductivity indicates the positive magnetoconductivity effect 
for the thin 1.2 nm Ta layer. 
Upon increasing the Ta layer thickness to 2.3\,nm, the Drude term decreases, and the optical conductivity changes become extended, involving the low-energy (around 2\,--\,3\,eV) and the high-energy (around 6\,--\,8\,eV) Lorentz bands. Upon further increasing the Ta layer thickness to 4.6\,nm, the Drude dc limit falls well below the weak localization limit, and the pronounced increase of the 6\,--\,8\,eV Lorentz band was discovered. The observed global band structure reconstruction may be indicative of the formation of a nearly localized many-body electron state \cite{Basko}.\\

See supplementary material for the atomic force microscopy (AFM) study of the MLFs and for details of the spectroscopic ellipsometry study.\\

This work was supported by the Grant 15-13778S of the Czech Science Foundation.\\

\end{document}